\documentclass[12pt,english]{paper}
\usepackage{amsthm,amsmath,amssymb}
\numberwithin{equation}{section}
\usepackage[T1]{fontenc}
\usepackage[latin1]{inputenc}
\usepackage{graphicx}
\usepackage{cite}
\usepackage{slashed}
\makeatletter

\newcommand{\bibstyle@aas}{\bibpunct{(}{)}{;}{a}{}{,}}
\makeatletter

\makeatletter

\usepackage{geometry}

\makeatletter

\usepackage{epsfig}

\makeatother

\makeatother

\makeatother

\usepackage{babel}
\makeatother
\begin{document}

\title{Measurement of Milli-Charged Particles with a moderately large cross section from the Earth's core at IceCube}

\author{Ye Xu}

\maketitle

\begin{flushleft}
School of Electronic, Electrical Engineering and Physics,  Fujian University of Technology, Fuzhou 350118, China
\par
e-mail address: xuy@fjut.edu.cn
\end{flushleft}

\begin{abstract}
It is assumed that heavy dark matter $\phi$ with O(TeV) mass captured by the Earth may decay to relativistic light milli-charged particles (MCPs). These MCPs could be measured by the IceCube neutrino telescope. The massless hidden photon model was taken for MCPs to interact with nuclei, so that the numbers and fluxes of expected MCPs may be evaluated at IceCube. Meanwhile, the numbers of expected neutrino background events were also evaluated at IceCube. Based on the assumption that no events are observed at IceCube in 10 years, the corresponding upper limits on MCP fluxes were calculated at 90\% C. L.. These results indicated that the MCPs from the Earth's core could be directly detected at O(1TeV) energies at IceCube when $2\times10^{-5}\lesssim\epsilon^2\lesssim4.5\times10^{-3}$. And a new region of 100 MeV < $m_{MCP}$ < 10 GeV and $4.47\times10^{-3}$ $\lesssim$ $\epsilon$ $\lesssim$ $9.41\times10^{-2}$ is ruled out in the $m_{MCP}$-$\epsilon$ plane with 10 years of IceCube data.
\end{abstract}
\begin{keywords}
Heavy dark matter, Milli-charged particles, Neutrino
\end{keywords}

\section{Introduction}
It is indicated by the Planck data with measurement of the cosmic microwave background that about 26\% of the overall energy density of the Universe is  non-baryonic dark matter (DM)\cite{Planck2016}. However, constraints on DM have become more and more stringent, and most of the parameter space for DM has been ruled out\cite{XENON1T,PANDAX,fermi,antares-icecube-dm-MW,icecubedm-sun,antaresdm-sun,CAST,GlueX,NGC1275,Chooz,dayabayMINOS}. The existence of a window for moderate interaction between nuclei and light DM candidates with masses below 100 GeV was emphasized in ref.\cite{ZF} (DM-nucleon cross section $\sim$ $\mu$b). A fraction of DM may make up of a particle species that interact much more strongly with SM particles than a typical DM with similar mass. Here a scenario where a small part of the DM interacts with the standard model (SM) particles with a moderately large cross section is assumed. In this scenario, a part of these particles with moderately large cross sections could reach the IceCube detectors and be measured by IceCube at O(TeV).
\par
The origin of DM is still unkoown, despite plenty of compelling observational evidence\cite{BHS}. In this DM scenario, there exist at least two DM species in the Universe. For example, O(TeV) DM and light DM particles(their masses $\ll$ 1 TeV). This heavy particle ($\phi$) is a thermal particle which is generated by the early universe. The bulk of present-day DM consists of them. The other is a stable light fermion($\chi$) which is the product of the decay of $\phi$ ($\phi\to\chi\bar{\chi}$), like the DM decay channel mentioned in Ref.\cite{FR}. This light fermion is a milli-charged particle (MCP), which is an alternative DM scenario\cite{GH,CY,FLN}, with a small electric charge $\epsilon e$ ($e$ is the electric charge for an electron and $\epsilon$<1). The searches for MCPs have been performed in cosmological and astrophysical observations, accelerator experiments, experiments for decay of ortho-positronium and Lamb shift, DM searches and so on, so that constraints on $\epsilon$ were determined by those observations\cite{CM,DHR,GH,DCB,SLAC,Xenon,LS,OP,Lamb,SUN}. Here the MCP mass is taken to be below 10 GeV. Due to the decay of long-living $\phi$ ($\tau_{\phi} > t_0$\cite{AMO,EIP}, $t_0\sim10^{17}$ s is the age of the Universe. Here $\tau_{\phi} \geq 10^{18}$ s), the present-day DM may also contain a very small component which is MCPs with energies of about $\displaystyle\frac{m_\phi}{2}$, where $m_{\phi}$ is the mass of $\phi$. In this scenario, besides, it is assumed that the decay of $\phi$ are only through $\phi\to\chi\bar{\chi}$.
\par
The $\phi$'s of the Galactic halo would be captured by the Earth when $\phi$ wind sweeps through the Earth. In this work, it is assumed that MCPs interact with nuclei with a moderately large cross section. Since the MCPs from the Sun's core hardly reach the Sun's surface ($\displaystyle\frac{R_s}{R_e}$ $\sim$ 1000, $R_e$ and $R_s$ are denoting the Earth's and Sun's radii, respectively), they can't be probed by the detectors on the Earth. One can only detect the high energy MCPs from the Earth's core (see Fig. 1). Those particles can be directly measured with the IceCube neutrino telescope via the deep inelastic scattering (DIS) with nuclei in the ice after they pass through the Earth and ice. The capability of the measurement of them will also be discussed in the present work. In this measurement, the background consists of neutrinos generated in cosmic ray interactions in the Earth's atmosphere and astrophysical neutrinos.
\section{Flux of MCPs which reach the Earth}
The $\phi$'s of the Galactic halo would collide with atomic nuclei in the Earth and be captured when their wind sweeps through the Earth. Those $\phi$'s inside the Earth can decay into MCPs at an appreciable rate. Then the number of those $\phi$'s is obtained in the way in Ref.\cite{BCH}
\begin{center}
\begin{equation}
\frac{dN}{dt}=C_{cap}-C_{ann}N^2-C_{evp}N
\end{equation}
\end{center}
where $C_{cap}$, $C_{ann}$ and $C_{evp}$ are the capture, the annihilation and the evaporation rates, respectively. The evaporation rate is only relevant when the DM mass < 5 GeV\cite{BCH}, which are much lower than my interested mass scale (m$_{\phi}$ $\sim$ O(TeV)). Thus their evaporation contributes to the accumulation in the Earth at a negligible level in the present work. The twice annihilation rate $\Gamma_{ann}=\displaystyle\frac{1}{2}C_{ann}N^2$. $\Gamma_{ann}$ is obtained by the following equation\cite{BCH,FST}
\par
\begin{center}
\begin{equation}
\Gamma_{ann}=\frac{C_{cap}}{2}tanh^2\left(\frac{t}{\tau}\right)\approx \frac{C_{cap}}{2} \quad with \quad t\gg\tau
\end{equation}
\end{center}
where $\tau=(C_{cap}C_{ann})^{-\frac{1}{2}}$ is a time-scale set by the competing processes of capture and annihilation. At late times $t\gg\tau$ one can approximate tanh$^2\displaystyle\frac{t}{\tau}$=1 in the case of the Earth\cite{BCH}. $C_{cap}$ is proportional to $\displaystyle\frac{\sigma_{\phi N}}{m_{\phi}}$\cite{BCH,JKK}, where $\sigma_{\phi N}$ is the scattering cross section between the nuclei and $\phi$'s. The spin-independent cross section is only considered in the capture rate calculation. Then $\sigma_{\phi N}$ is taken to be 10$^{-44}$ cm$^2$ for $m_{\phi} \sim$ O(TeV) \cite{XENON1T,PANDAX}. Besides, one knows that $\phi$'s are concentrated around the center of the Earth from Ref.\cite{FST}.
\par
The MCPs which reach the IceCube detector are produced by the decay of $\phi$'s in the Earth's core. Those MCPs have to pass through the Earth and interact with nuclei inside the Earth. Then the number N$_e$ of MCPs produced in the Earth's core is obtained by the following equation:
\begin{center}
\begin{equation}
\begin{aligned}
N_e &=2N_0\left(exp(-\frac{t_0}{\tau_{\phi}})-exp(-\frac{t_0+T}{\tau_{\phi}})\right) \qquad with \quad T \ll \tau_{\phi}\\
    &\approx 2N_0\frac{T}{\tau_{\phi}}exp(-\frac{t_0}{\tau_{\phi}})
\end{aligned}
\end{equation}
\end{center}
where N$_0$=$\displaystyle\int^{t_e}_0 \displaystyle\frac{dN}{dt} dt$ is the number of $\phi$'s captured in the Earth. t$_e$ and t$_0$ are the ages of the Earth and the Universe, respectively. T is the lifetime of taking data for IceCube and taken to be 10 years. $R_e$ is denoting the Earth's radius.
\par
Then the flux $\Phi_{MCP}$ of MCPs, which reach the IceCube detector, from the Earth's core is described by
\begin{center}
\begin{equation}
\Phi_{MCP}=\frac{N_e}{4\pi R_e^2}exp(-\displaystyle\frac{R_e}{L^{\chi}_{e}})
\end{equation}
\end{center}
where $L^{\chi}_e$ is denoting the MCP interaction length with the Earth.
\section{MCP and neutrino interactions with nuclei}
A second unbroken "mirror" U(1)$^{\prime}$ was introduced in the hidden photon model. The corresponding massless hidden photon field may have a kinetic mixing to the SM photon, so that a MCP under U(1)$^{\prime}$ appears to have a small coupling to the SM photon\cite{Holdom}. $\epsilon$ is the kinetic mixing parameter between those two kinds of photons. This model is taken for MCPs to interact with nuclei via a neural current (NC) interaction mediated by the mediator generated by the kinetic mixing between the SM and massless hidden photons. There is only a well-motivated interaction allowed by SM symmetries that provide a "portal" from the SM particles into the MCPs. This portal is $\displaystyle\frac{\epsilon}{2}F_{\mu\nu}F^{\prime\mu\nu}$. Then its interaction Lagrangian can be written as follows:
\begin{center}
\begin{equation}
\mathcal{L} =\sum_qe_q\bar{q}\gamma^{\mu}qA_{\mu} -\frac{1}{4}F^{\prime}_{\mu\nu}F^{\prime\mu\nu}+\bar{\chi}(i\slashed{D}-m_{\chi})\chi-\frac{\epsilon}{2}F_{\mu\nu}F^{\prime\mu\nu}
\end{equation}
\end{center}
where the sum runs over quark flavors in the nucleon and $e_q$ is the electric charge of the quark. $A_{\mu}$ is the vector potential of the SM photon. $F^{\prime}_{\mu\nu}$, $F_{\mu\nu}$ are the field strength tensor of the hidden and SM photons, respectively. $m_{\chi}$ is the MCP's mass.
\par
The DIS cross section of MCPs on nuclei is computed with the same model in Sec. 3 in my previous work\cite{SUN}. Since the MCP-mediator coupling is equal to $\epsilon^2\alpha$, the DIS cross section of MCPs on nuclei is equivalent to $\epsilon^2$ times as much as that of electrons on nuclei via a NC interaction under electromagnetism, that is
\begin{center}
\begin{equation}
\sigma_{\chi N}\approx\epsilon^2\sigma^{\gamma}_{eN}
\end{equation}
\end{center}
where $\chi$ denotes a MCP with $\epsilon e$, N is a nucleon. $\sigma^{\gamma}_{eN}$ is the cross section depending on $\gamma$ exchange between elections and nuclei.
The total DIS cross sections of MCPs on nuclei may be approximately expressed as a simple power-law form in the energy range 1 TeV-1 PeV\cite{SUN}
\begin{center}
\begin{equation}
\sigma_{\chi N}\approx1.756\times10^{-31}\epsilon^2 cm^2 \left(\frac{E_{\chi}}{1GeV}\right)^{0.179}
\end{equation}
\end{center}
where E$_{\chi}$ is the MCP energy.
\par
The DIS cross-section for neutrino interaction with nuclei is computed in the lab-frame and given by simple power-law forms\cite{BHM} for neutrino energies above 1 TeV:
\begin{center}
\begin{equation}
\sigma_{\nu N}(CC)=4.74\times10^{-35} cm^2 \left(\frac{E_{\nu}}{1 GeV}\right)^{0.251}
\end{equation}
\end{center}
\par
\begin{center}
\begin{equation}
\sigma_{\nu N}(NC)=1.80\times10^{-35} cm^2 \left(\frac{E_{\nu}}{1 GeV}\right)^{0.256}
\end{equation}
\end{center}
where $\sigma_{\nu N}(CC)$ and $\sigma_{\nu N}(NC)$ are the DIS cross sections for neutrino scattering on nuclei via the charge current (CC) and neutral current (NC) interactions, respectively. $E_{\nu}$ is the neutrino energy.
\par
The MCP and neutrino interaction lengths can be obtained by
\begin{center}
\begin{equation}
L^{\nu,\chi}=\frac{1}{N_A\rho\sigma_{\nu,\chi N}}
\end{equation}
\end{center}
where $N_A$ is the Avogadro constant, and $\rho$ is the density of matter, which MCPs and neutrinos interact with.
\section{Evaluation of the numbers of expected MCPs and neutrinos at IceCube}
The IceCube detector is deployed in the deep ice below the geographic South Pole\cite{icecube2004}. The high energy MCPs which are passing through the IceCube detector would interact with the nuclei inside IceCube. This is similar to the NC DIS of neutrino interaction with nuclei, whose secondary particles would develop into a cascade at IceCube.
\par
The number N$_{det}$ of expected MCPs obeys the following equation:
\begin{center}
\begin{equation}
\frac{dN_{det}}{dE} =C_1\times C_2\times A_{eff}(E)\Phi_{MCP} P(E)
\end{equation}
\end{center}
where $A_{eff}(E)$ obtained from the Fig. 2 in Ref.\cite{icecube2014a} is denoting the effective observational area for IceCube. E is denoting the energy of an incident particle. $C_1$ is equal to 68.3\% (that is 68.3\% of the MCP events reconstructed with IceCube fall into a window caused by one standard energy uncertainty). $C_2$ is equal to 50\% (that is 50\% of the MCP events reconstructed with IceCube fall into a window caused by one median angular uncertainty). $P(E)$ can be given by the following equation:
\begin{center}
\begin{equation}
P(E)=1-exp(-\displaystyle\frac{D}{L^{\chi}_{ice}}).
\end{equation}
\end{center}
where $L^{\chi}_{ice}$ is denoting the MCP interaction length with the ice. D is denoting the effective length in the IceCube detector and taken to be 1 km in this work.
\par
After rejecting track-like events, the background remains two sources: astrophysical and atmospheric neutrinos which pass through the detector of IceCube. Only a NC interaction with nuclei is relevant to muon neutrinos considered here. The astrophysical neutrinos flux can be described by\cite{icrc2023}
\begin{center}
\begin{equation}
\Phi_{\nu}^{astro}=\Phi_{astro}\times\left(\displaystyle\frac{E_{\nu}}{100TeV}\right)^{-(\alpha+\beta log_{10}(\frac{E_{\nu}}{100TeV}))}\times10^{-18}GeV^{-1} cm^{-2}s^{-1}sr^{-1}
\end{equation}
\end{center}
where $\Phi_{\nu}^{astro}$ is denoting the total astrophysical neutrino flux. The coefficients, $\Phi_{astro}$, $\alpha$ and $\beta$ are given in Tab. 2 in Ref.\cite{icrc2023}. The atmospheric neutrinos flux can be described by\cite{SMS}
\begin{center}
\begin{equation}
\Phi_{\nu}^{atm} = C_{\nu}\left(\displaystyle\frac{E_{\nu}}{1GeV}\right)^{-(\gamma_0+\gamma_1x+\gamma_2x^2)}GeV^{-1} cm^{-2}s^{-1}sr^{-1}
\end{equation}
\end{center}
where $x=log_{10}(E_{\nu}/1GeV)$. $\Phi_{\nu}^{atm}$ is denoting the atmospheric neutrino flux. The coefficients, $C_{\nu}$ ($\gamma_0$, $\gamma_1$ and $\gamma_2$) are given in Table III in Ref.\cite{SMS}.
\par
The neutrinos fallen into the energy and angular windows mentioned above would also be regarded as signal candidate events, so the evaluation of the number of expected neutrinos has to be performed by integrating over the region caused by these windows. Then the number of expected neutrinos N$_{\nu}$ obeys the following equation:
\begin{center}
\begin{equation}
\frac{dN_{\nu}}{dE} = \int_T \int_{\theta_{min}}^{\theta_{max}} A_{eff}(E)(\Phi_{\nu}^{astro}+\Phi_{\nu}^{atm}) P(E,\theta)\frac{2\pi R_e^2 sin2\theta}{D_e(\theta)^2} d\theta dt
\end{equation}
\end{center}
where $D_e(\theta)=2R_e cos\theta$ is denoting the distance through the Earth. $\theta$ is the zenith angle at IceCube (see Fig. 1). $\theta_{min}$ = 0 and $\theta_{max}$ = $\sigma_{\theta}$. $\sigma_{\theta}$ is denoting the median angular uncertainty for cascades at IceCube. The standard energy and median angular uncertainties can be obtained from the Ref.\cite{icecube2021ICRC} and Ref.\cite{icecube2013}, respectively. $P(E,\theta)$ can be given by
\begin{center}
\begin{equation}
P(E,\theta)=exp(-\displaystyle\frac{D_e(\theta)}{L^{\nu}_{e}})\left(1-exp(-\displaystyle\frac{D}{L^{\nu}_{ice}})\right)
\end{equation}
\end{center}
where $L^{\nu}_{e,ice}$ are denoting the neutrino interaction length with the Earth and the ice, respectively.
\section{Results}
The distributions and numbers of expected MCPs and neutrinos were evaluated in the energy range 1 TeV-1 PeV assuming 10 years of IceCube data. Fig. 2 shows the distributions with an energy bin of 100 GeV of expected MCPs and neutrinos. Compared to MCPs with $\epsilon^2$=$4.5\times10^{-3}$ and $\tau_{\phi} = 10^{18}$ s (the magenta dash dot line), the numbers of neutrino events per energy bin are at least smaller by 3 orders of magnitude in the energy range 1 TeV-1 PeV. As shown in Fig. 2, the dominant background is caused by atmospheric neutrinos at energies below 200 TeV but astrophysical neutrinos at energies above about 400 TeV in this measurement.
\par
The numbers of expected neutrinos (see black solid line) are shown in Fig. 3. The evaluation of the numbers of expected neutrinos was performed through integrating over the region caused by the energy and angular uncertainties described above. The black dash line denotes the number of expected atmospheric neutrinos. This figure indicates the neutrino background is unable to be ignored in the interested energy range in this measurement (for example, the number of expected neutrinos is about 0.6 at 1 TeV). The numbers of expected MCPs with $\epsilon^2=9\times10^{-4}$ and $\tau_{\phi} = 10^{18}$ s can reach about 19 and 1 at 1 TeV and 12 TeV at IceCube, respectively, as shown in Fig. 3 (see the red dash line). Fig. 3 also presents the MCPs with $\epsilon^2=2\times10^{-5}$ (see the blue dot line) and $\epsilon^2=4.5\times10^{-3}$(see the magenta dash dot line) could be detected below about 2 TeV and 1 TeV at IceCube, respectively, when $\tau_{\phi} = 10^{18}$ s. With $\epsilon^2$ of about $9\times10^{-4}$, as shown in Fig. 3, the number of expected MCPs would reach a maximum value at IceCube.
\par
Ref.\cite{icecube2024} presents a all-sky search for transient astrophysical neutrino emission with 10 years of IceCube cascade events. This analysis has not found any significant indication of neutrinos from the Earth's core. Since the MCP and neutrino signals are hard to distinguish at IceCube, it is a reasonable assumption that no events are observed in this search for MCPs due to the decay of $\phi$ in the Earth's core at IceCube in 10 years. The corresponding upper limit on MCP flux at 90\% C.L. was calculated with the Feldman-Cousins approach\cite{FC} (see the black solid line in Fig. 4). Fig. 4 also presents the fluxes of expected MCPs with $\epsilon^2=2\times10^{-5}$ (blue dot line),$9\times10^{-4}$ (red dash line) and $4.5\times10^{-3}$ (magenta dash dot line). That limit excludes the MCP fluxes with $\epsilon^2=9\times10^{-4}$ below about 5.3 TeV.
\section{Discussion and Conclusion}
With $\epsilon^2$ = $9\times10^{-4}$, hence, the MCPs from the Earth's core can be measured in the energy ranges 5.3-12 TeV at IceCube when $\tau_{\phi} = 10^{18}$ s. With $\epsilon^2$ = $2\times10^{-5}$ and $4.5\times10^{-3}$, the ones from the Earth's core can also be probed at IceCube. Based on the results described above, it is a reasonable conclusion that those MCPs could be directly detected  at O(1TeV) at IceCube when $2\times10^{-5}\lesssim\epsilon^2\lesssim4.5\times10^{-3}$. Since these constraints are only given by the assumptions mentioned above, certainly, the experimental collaborations, like the IceCube collaboration, should be encouraged to conduct an unbiased analysis with the data of IceCube.
\par
Since $\Phi_{MCP}$ is roughly proportional to $\displaystyle\frac{1}{\tau_{\phi}}$ (see Eqn. (2.3)), the above results actually depends on the lifetime of heavy DM, $\tau_{\phi}$. If $\tau_{\phi}$ =$3\times10^{19}$ s, for example, the numbers of expected MCPs are about 20 times less than that with $\tau_{\phi}=10^{18}$ s at IceCube.
\par
Likewise, the upper limit for $\epsilon^2$ at 90\% C.L. can be calculated with the Feldman-Cousins approach. Fig. 5 shows these limits with $\tau_{\phi}$ = $10^{18}$ s (see red solid line), $7\times10^{18}$ s (see blue dash line) and $3\times10^{19}$ s (see magenta dot line), respectively. If $m_{\phi}$ = 2 TeV (the corresponding MCP energy is just 1 TeV), as shown in Fig. 5, the region of $\epsilon^2<8.86\times10^{-3}$ (that is $\epsilon$ <$9.41\times10^{-2}$) is ruled out when $\tau_{\phi}=10^{18}$ s.
\par
The MCP mass, $m_{\chi}$, is at least taken to be less than 10 GeV, as mentioned in Sec. 1. Then the region of $\epsilon < 9.41\times10^{-2}$ is ruled out at 90\% C.L. in the $m_{\chi}$-$\epsilon$ plane, when $m_{\chi} < $ 10 GeV. This result is shown in Fig. 6. Considering the capability of the measurement of those MCPs, besides, we had to constrain on $\epsilon$ ($\epsilon^2\gtrsim2\times10^{-5}$, that is $\epsilon\gtrsim4.47\times10^{-3}$). To compare to other observations on MCPs, this figure also shows the $\epsilon$ bounds from cosmological and astrophysical observations\cite{CM,DHR,DGR,JR}, accelerator and fixed-target experiments\cite{DCB,SLAC}, experiments for decay of ortho-positronium\cite{OP}, Lamb shift\cite{Lamb} and measurement of MCPs from the Sun's core\cite{SUN}. A new region of 100 MeV < $m_{\chi}$ < 10 GeV and $4.47\times10^{-3}$ < $\epsilon$ $\lesssim$ $9.41\times10^{-2}$ is ruled out in the $m_{\chi}$-$\epsilon$ plane with 10 year of IceCube data, as shown in Fig. 6.
\par
Since the decay of $\phi$'s into MCPs can lead to extra energy injection during recombination and reionization eras in the early universe, the parameters in this DM scenario may be constrained by early universe observations. Since the $\phi$ lifetime is greater than the age of the Universe, however, $\Omega_{MCPs} h^2 \lesssim 10^{-11}\Omega_{DM}h^2$ in this scenario. Then this result is consistent with that in Ref.\cite{DDRT}, since Ref.\cite{DDRT} presented that the cosmological abundance of MCPs was strongly constrained by the Planck data, that was $\Omega_{MCPs}h^2<0.001$.
\section{Acknowledgements}
This work was supported by the National Natural Science Foundation
of China (NSFC) under the contract No. 11235006, the Science Fund of
Fujian University of Technology under the contracts No. GY-Z14061 and GY-Z13114 and the Natural Science Foundation of
Fujian Province in China under the contract No. 2015J01577.
\par

\newpage

\begin{figure}
 \centering
 \includegraphics[width=0.9\textwidth]{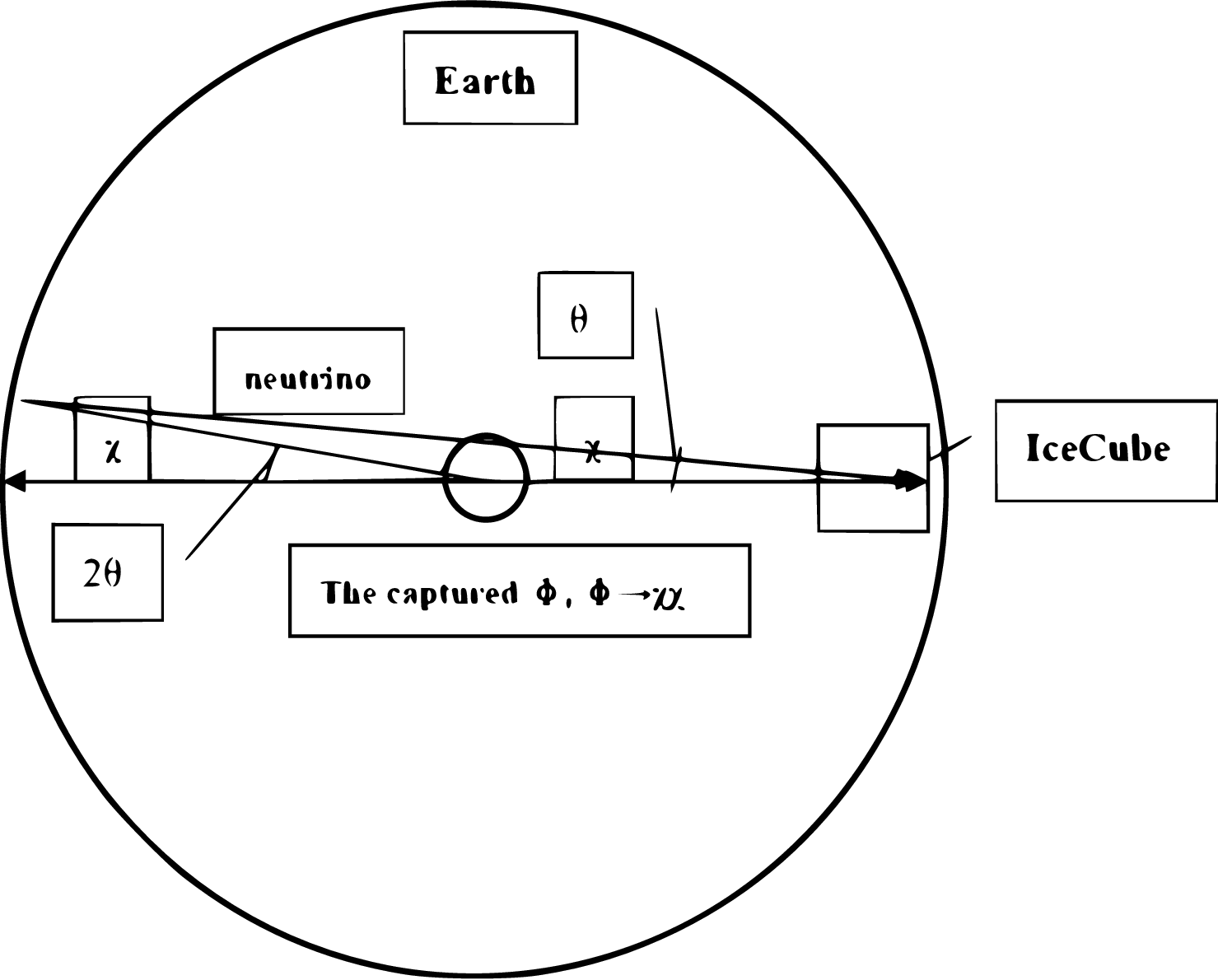}
%%% bb = left_bottom_X, left_bottom_Y, right_top_X, right_top_Y
%%% scale through "set width"
 \caption{MCPs, due to the decay of heavy DM particles captured in the earth core, pass through the Earth and ice and could be measured with the detector like IceCube neutrino telescope}
 \label{fig:fig}
\end{figure}

\begin{figure}
 \centering
 \includegraphics[width=0.9\textwidth]{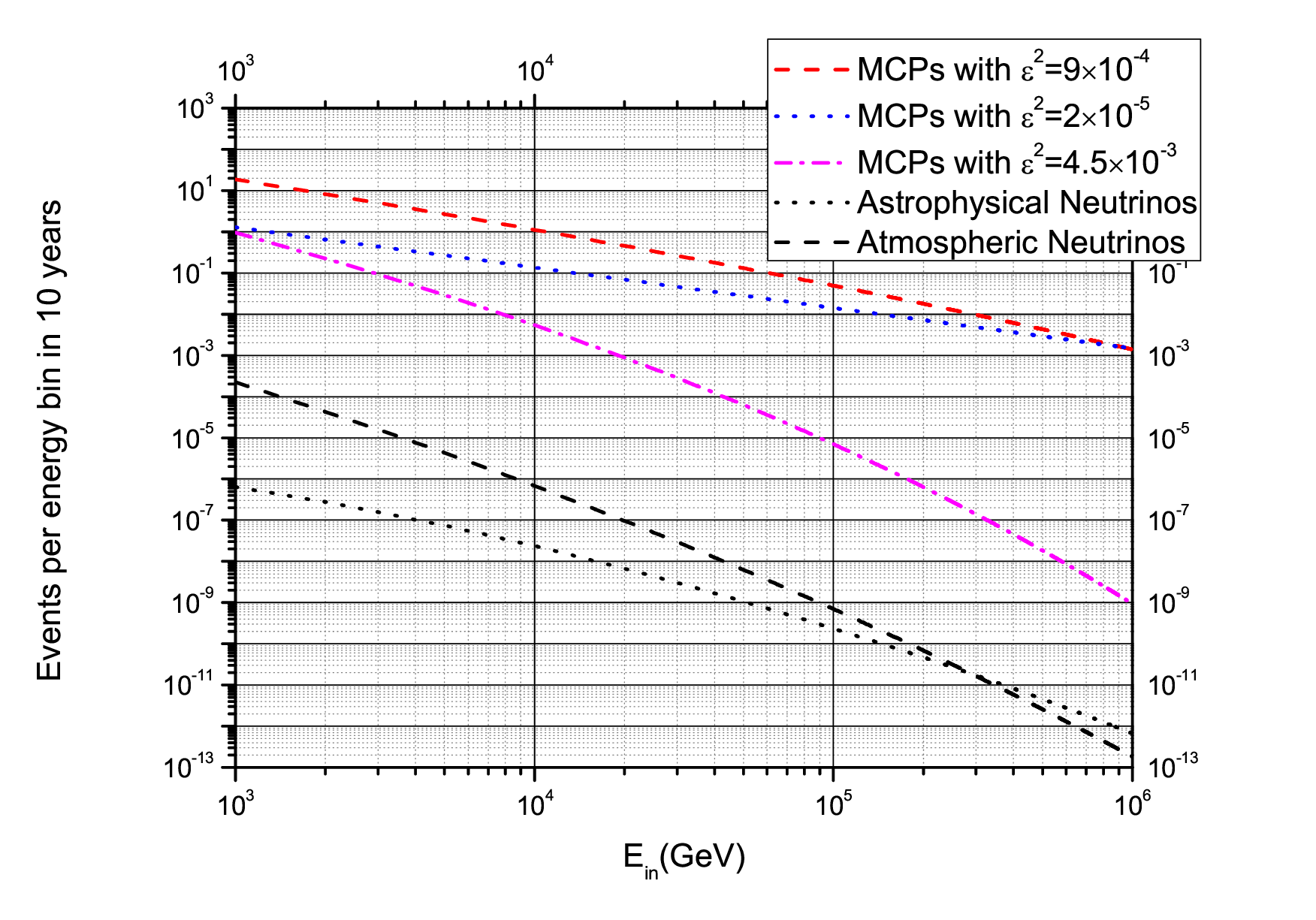}
%%% bb = left_bottom_X, left_bottom_Y, right_top_X, right_top_Y
%%% scale through "set width"
 \caption{Distributions of expected MCPs with $\tau_{\phi}$ = $10^{18}$ s and $\epsilon^2=2\times10^{-5}$, $9\times10^{-4}$, $4.5\times10^{-3}$ and astrophysical and atmospheric neutrinos. Their energy bins are 100 GeV.}
 \label{fig:E_bin}
\end{figure}

\begin{figure}
 \centering
 %\includegraphics[bb=0 0 200 300, width=3.8cm]{energy}
 %\hspace{0.7\textwidth}
 \includegraphics[width=0.9\textwidth]{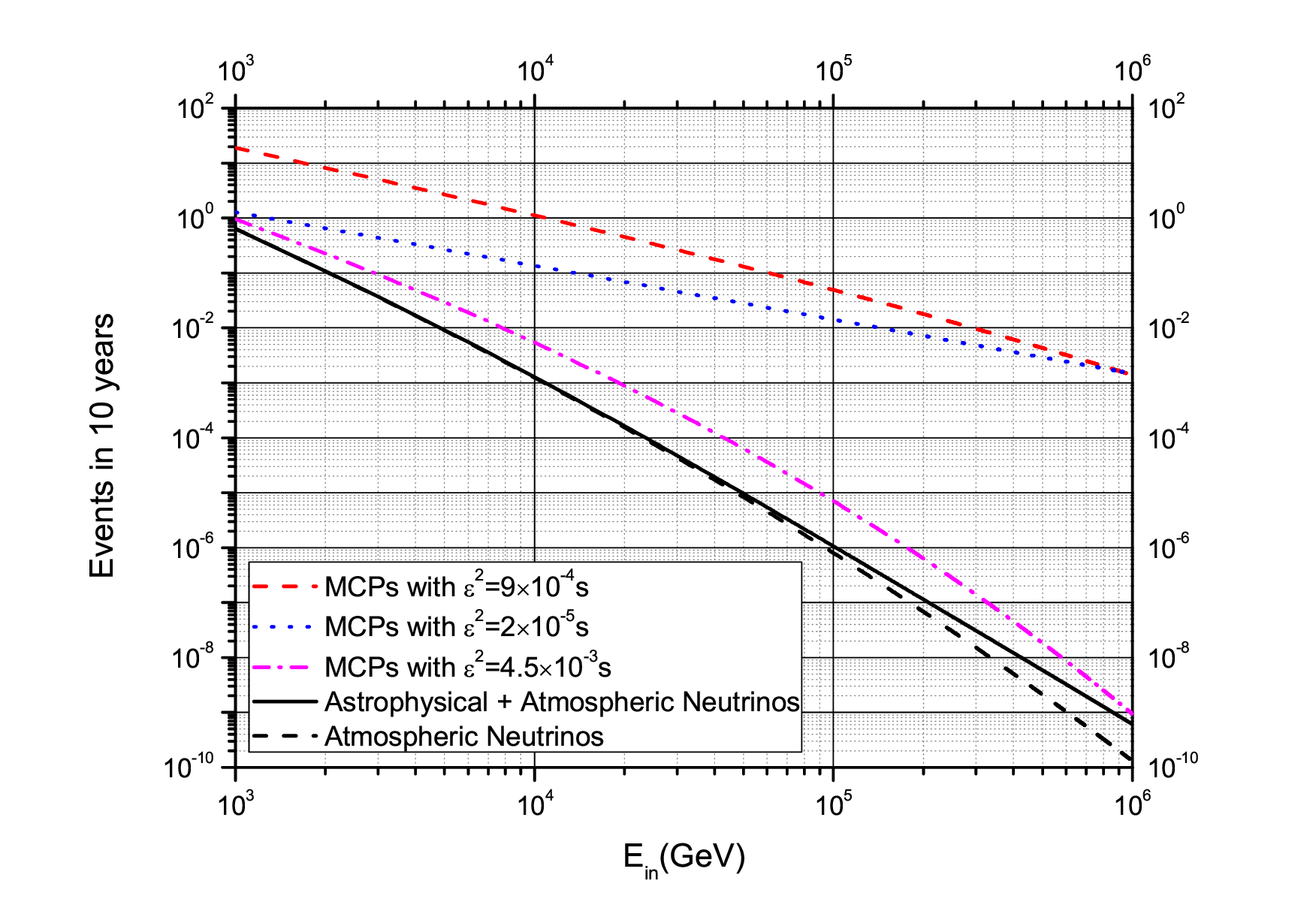}
 \caption{With the different $\epsilon^2$ (= $2\times10^{-5}$, $9\times10^{-4}$ and $4.5\times10^{-3}$) and $\tau_{\phi}$ = $10^{18}$, the numbers of expected MCPs were evaluated assuming 10 years of IceCube data, respectively. The evaluation of numbers of expected neutrinos was performed by integrating over the region caused by one standard energy and median angular uncertainties.}
 \label{fig:events_1e18}
\end{figure}

\begin{figure}
 \centering
 \includegraphics[width=0.9\textwidth]{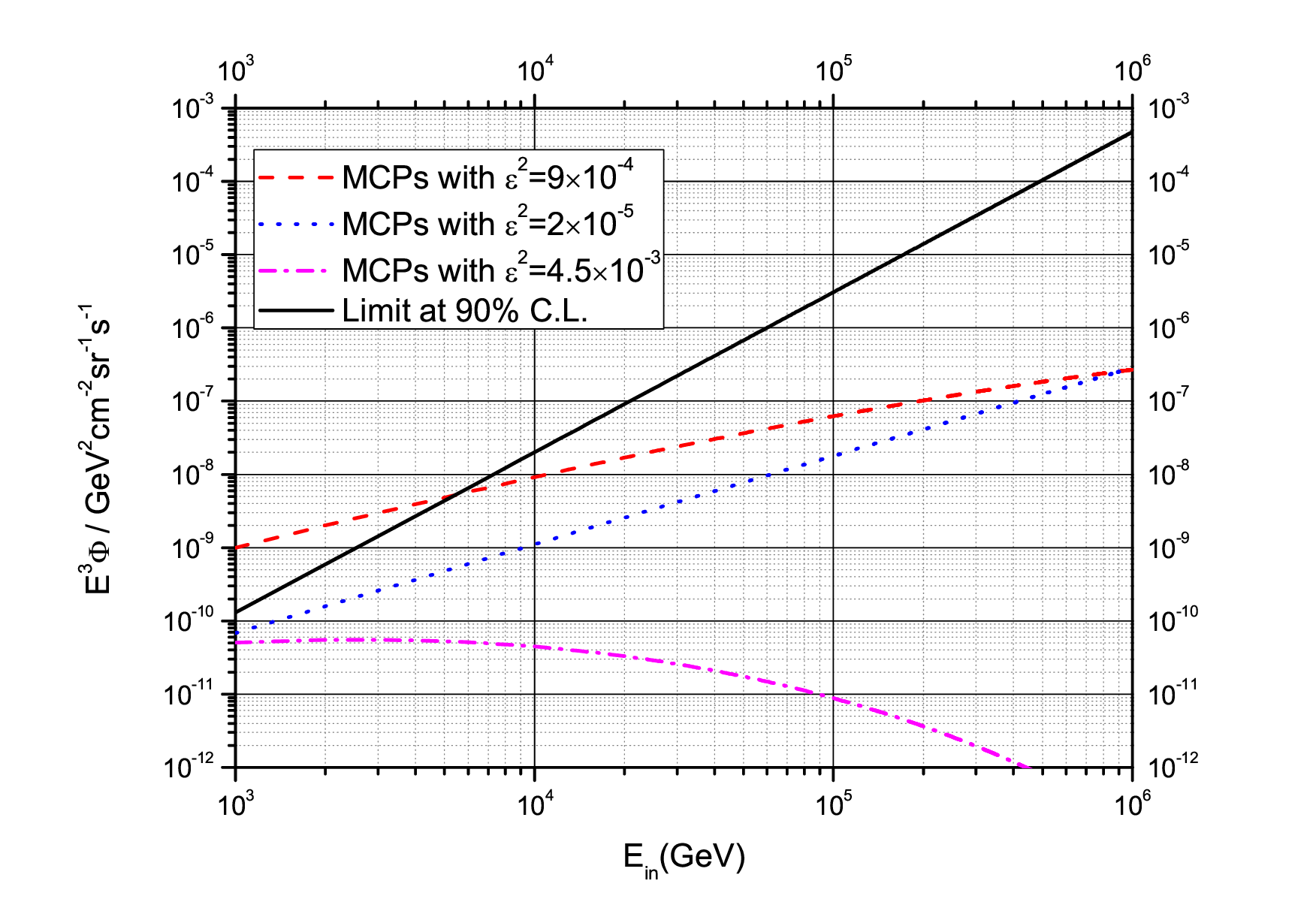}
%%% bb = left_bottom_X, left_bottom_Y, right_top_X, right_top_Y
%%% scale through "set width"
 \caption{With the different $\epsilon^2$ (= $2\times10^{-5}$, $9\times10^{-4}$ and $4.5\times10^{-3}$) and $\tau_{\phi}$ = $10^{18}$, the fluxes of expected MCPs were estimated at IceCube, respectively. Assuming no observation at IceCube in 10 years, the upper limit at 90\% C.L. was also computed.}
 \label{fig:flux_1e18}
\end{figure}

\begin{figure}
 \centering
 %\includegraphics[bb=0 0 200 300, width=3.8cm]{energy}
 %\hspace{0.7\textwidth}
 \includegraphics[width=0.9\textwidth]{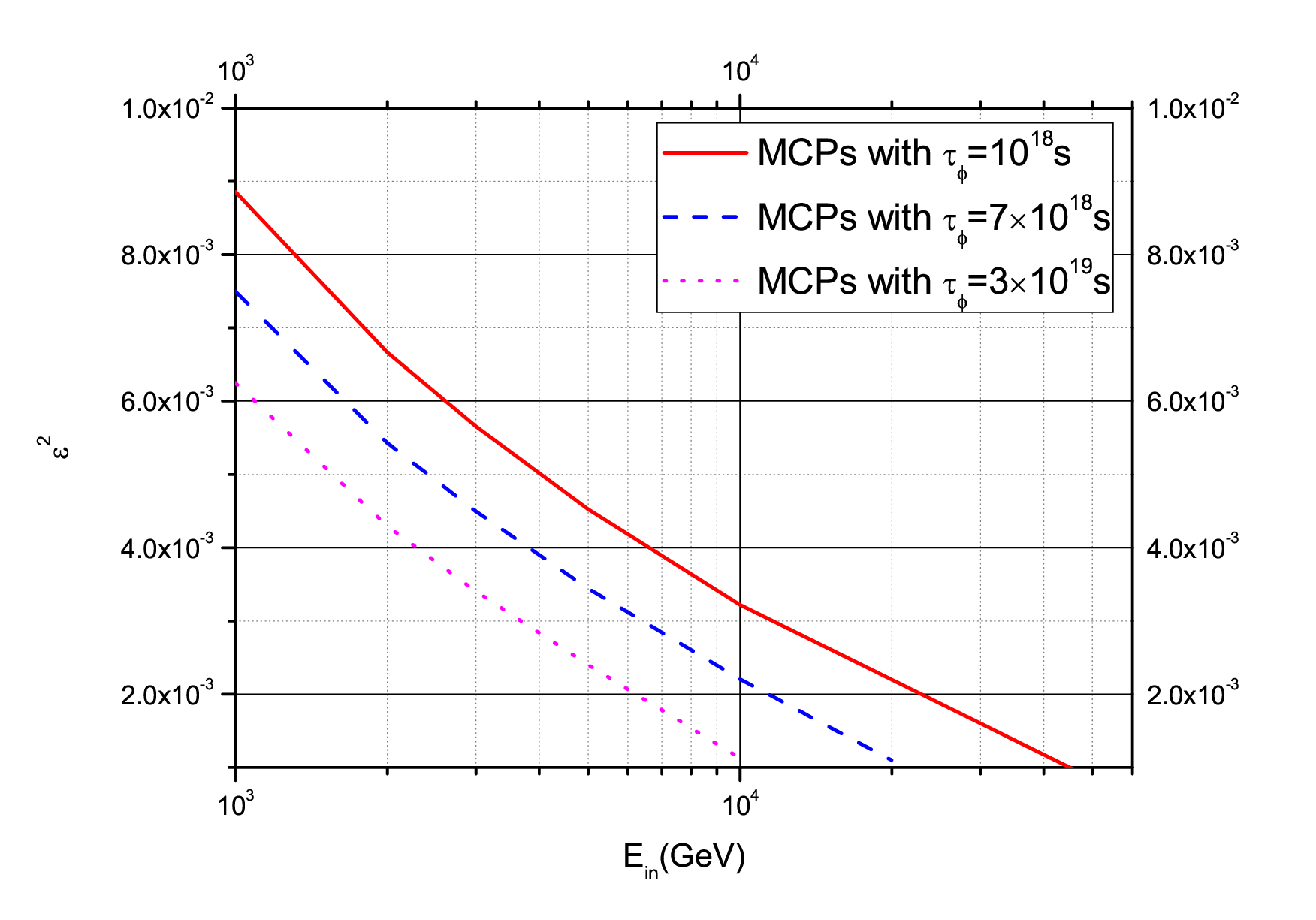}
 \caption{With the different $\tau_{\phi}$ (= $10^{18}$ s, $7\times10^{18}$ s and $3\times10^{19}$ s), the upper limit on $\epsilon$ at 90\% C.L. was computed, respectively, assuming no observation at IceCube in 10 years.}
 \label{fig:uplimit_epsilon2}
\end{figure}

\begin{figure}
 \centering
 %\includegraphics[bb=0 0 200 300, width=3.8cm]{energy}
 %\hspace{0.7\textwidth}
 \includegraphics[width=0.9\textwidth]{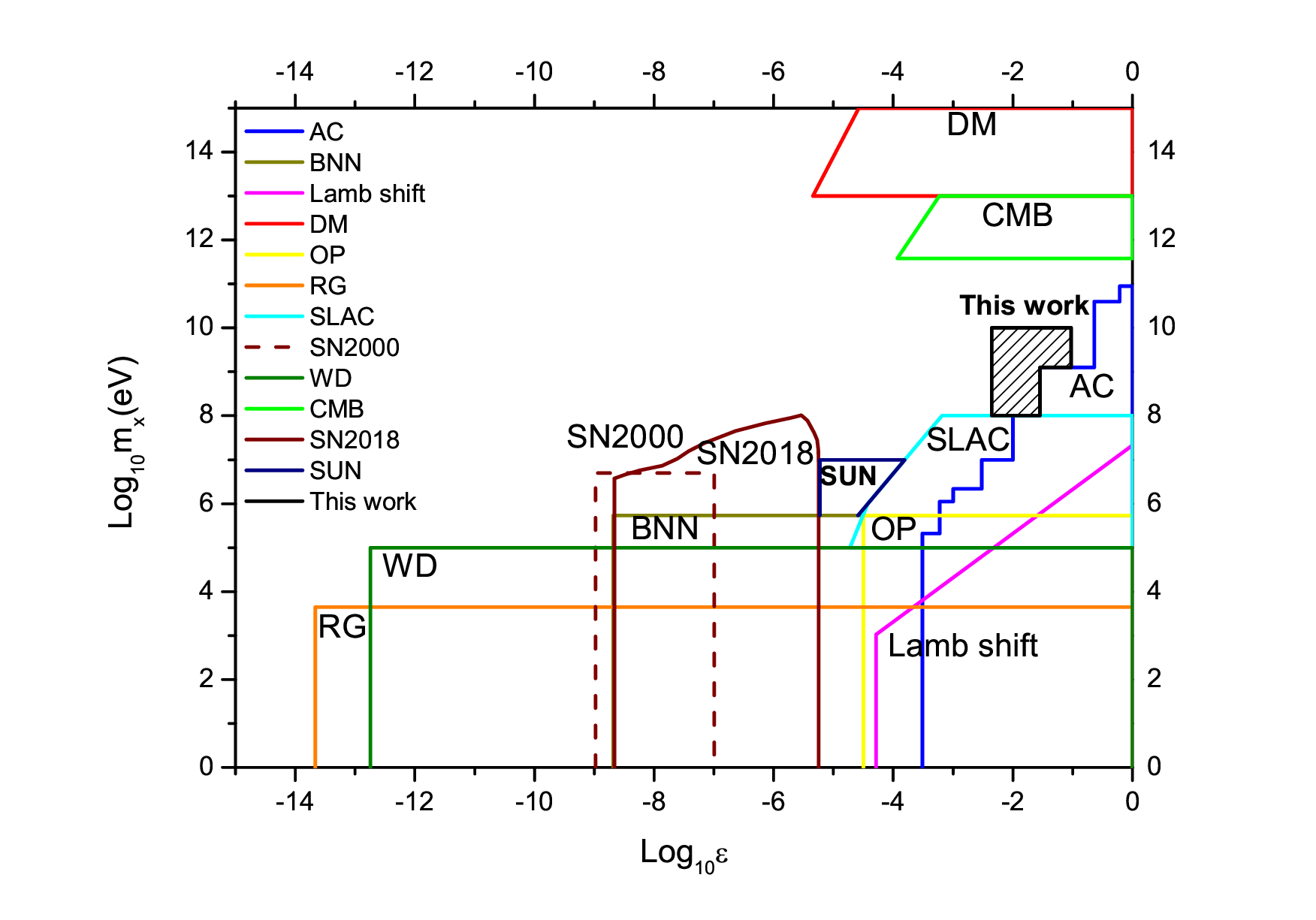}
 \caption{If $m_{\phi}$=2 TeV, a new region (shaded region) is ruled out in the $m_{\chi}$ vs. $\epsilon$ plane, when $\tau_{\phi}=10^{18}$ s. Meanwhile, the bounds from plasmon decay in red giants (RG)\cite{DHR}, plasmon decay in white dwarfs (WD)\cite{DHR}, cooling of the Supernova 1987A (SN2000\cite{DHR}, SN2018\cite{CM}), accelerator (AC)\cite{DCB} and fixed-target experiments (SLAC)\cite{SLAC}, the Tokyo search for the invisible decay of ortho-positronium (OP)\cite{OP}, the Lamb shift\cite{Lamb}, big bang nucleosynthesis (BBN)\cite{DHR}, cosmic microwave background (CMB)\cite{DGR}, dark matter searches (DM)\cite{JR} and measurement of MCPs from the sun's core\cite{SUN} are also plotted on this figure.}
 \label{fig:epsilon_bound}
\end{figure}

\end{document}